\documentclass{appolb}%
\usepackage{epsfig}
\usepackage{rotating}
\usepackage{hyperref}
\usepackage{comment,color}
\usepackage{braket}
\usepackage{amssymb,amsmath,bm}
\usepackage{setspace,lipsum}
\usepackage{amsmath}
\usepackage{amsfonts}
\usepackage{amssymb}
\usepackage{graphicx}%
\providecommand{\U}[1]{\protect\rule{.1in}{.1in}}
\begin{document}

\title{Modelling the inverse Zeno effect for the neutron decay}
\author{Francesco~Giacosa
\address{Institute of Physics, Jan Kochanowski University, 25-406 Kielce, Poland}
}
\maketitle

\begin{abstract}
Beam and trap methods find incompatible results for the lifetime of the
neutron: the former delivers a value which is about $8.7\pm2.1$ s longer than
the latter. Very recently (1906.10024) it has been proposed that the inverse
Zeno effect (IZE) could be responsible for the shorter lifetime in trap
experiments. Here, we compare two different models of measurement, one
obtained by bang-bang measurements and by a continuous measurement: the IZE
turns out to be in both cases very similar, showing that the results do not
depend on the details of the measurement process.

\end{abstract}


\textit{Introductory remarks:} The lifetime of the neutron represents an
unsolved puzzle: different measurement types find incompatible results
\cite{wietfeldt}. On the one hand, experiments based on the so-called beam
method -in which protons emitted by a neutrons beam are detected- lead to the
average $\tau_{\mathrm{n}}^{\text{beam}}=888.1\pm2.0$ s; on the other hand,
trap or cavity experiments, in which surviving neutrons confined in a magnetic
or gravitational trap are counted, deliver the result $\tau_{\mathrm{n}%
}^{\text{trap}}=879.37\pm0.58$ s. The deviation $8.7\pm2.1$ s s corresponds to
a $4\sigma$ discrepancy.

In recent works, this mismatch has been interpreted \cite{fornal,berezhiani}
in the following way: an invisible dark decay of the neutron causes the beam
method (which detects the emitted protons) to erroneously measure a longer
lifetime. In this context, the correct lifetime is given by $\tau_{\mathrm{n}%
}^{\text{trap}}$. This approach involves beyond-standard-model physics and
could be problematic for what concerns the stability of neutron stars
\cite{baym} as well as the present knowledge of neutron decay parameters
\cite{dubbers}.

Also very recently, an alternative idea was discussed in Ref. \cite{izen}.
Here, the inverse Zeno effect (IZE) increases the decay rate in trap
experiments, where about $10^{8}$ neutrons are kept in a cold environment.
Namely, the dephasing/decoherence of the neutron state may occur sufficiently
fast to reach the regime in which the IZE is realized. Thus, within this idea,
the correct lifetime is given by $\tau_{\mathrm{n}}^{\text{beam}}$.

In this work, we compare two different models of measurements leading to the
IZE. Besides the model based on continuous measurement used in Ref.
\cite{izen}, we also test the model based on instantaneous bang-bang
measurements: qualitatively similar results are obtained, thus confirming that
the IZE is not dependent on the details of the model.

\textit{Different realizations of the IZE:} The decay law in quantum mechanics
is not exactly an exponential \cite{fonda} (see Refs. \cite{nonexpqft} for the
analogous result in quantum field theory). If intermediate measurements are
performed, the decay width can change \cite{dega}, and eventually be sizably
reduced (QZE, \cite{sudarshan}). In some conditions, however, also the IZE
(larger measured decay width) is possible \cite{kk,fp,koshino}.

Let us consider a certain decay width, parametrized by the function
$\Gamma(\omega),$ where $\omega$ is the energy of the final state minus the
energy in the initial state (it then formally ranges between $0$ and $\infty
$). The usual decay width is obtained by setting $\omega=\omega
_{\text{onshell}},$ $\Gamma_{\text{onshell}}=\Gamma(\omega_{\text{onshell}}).$
The form of $\Gamma(\omega)$ depends on the details of the unstable state, but
is zero for $\omega<0$ and for $\omega\rightarrow\infty$. As shown in\ Ref.
\cite{kk,fp} (see also the review \cite{koshino}) the measured decay widths
read%
\begin{equation}
\Gamma^{\text{meas}}(\tau,\omega_{C})=\int_{0}^{\omega_{C}}f(\tau
,\omega)\Gamma(\omega)d\omega\text{ ,}%
\end{equation}
where $\omega_{C}$ is the maximal off-shellness, typically linked to the
formation process of the unstable state. The response function $f(\tau
,\omega)$ models the measurement, which averagely takes place at $\tau,$
$2\tau,$ etc. The three general properties of $f$ are: $\int_{-\infty}%
^{\infty}f(\tau,\omega)d\omega=1$ , $f(\tau\rightarrow\infty,\omega
)=\delta(\omega-\omega_{\text{onshell}})$ , $f(\tau\rightarrow0,\omega)=$
small const. The first is the normalization, the second implies that, for an
undisturbed system, the `on-shell' decay is obtained, the third implies that
$\Gamma^{\text{meas}}(\tau\rightarrow0)=0$ (QZE). The details of the function
$f(\tau,\omega)$ depend on which type of measurement is performed. For
instantaneous ideal measurements performed at times $\tau,$ $2\tau,...$ and
for a continuous measurement of the final state, it reads respectively
\cite{kk,fp}
\begin{equation}
f_{1}(\tau,\omega)=\frac{\tau}{2\pi}\text{\textrm{sinc}}^{2}\left[  \left(
\omega-\omega_{\text{onshell}}\right)  \frac{\tau}{2}\right]  \text{ , }%
f_{2}(\tau,\omega)=\frac{1}{\pi\tau}\left[  \left(  \omega-\omega
_{\text{onshell}}\right)  ^{2}+\tau^{-2}\right]  ^{-1}\text{.}%
\end{equation}
For each choice of $f$ one has a different measured decay width, $\Gamma
_{k}^{\text{meas}}(\tau,\omega_{C})=\int_{0}^{\omega_{C}}f_{k}(\tau
,\omega)\Gamma(\omega)d\omega.$ More complicated measurement models would lead
to different response function (for the possible effect of imperfect
measurements, see Ref. \cite{giacosapra}). The QZE is realized when
$\Gamma^{\text{meas}}<\Gamma_{\text{onshell}},$ while the IZE when
$\Gamma^{\text{meas}}>\Gamma_{\text{onshell}}$. Both the QZE and the IZE have
been experimentally verified on a genuinely unstable quantum system
\cite{reizen2}.\ 

Next, we turn to the specific case that we are interested in: the weak decay
of the neutron, for which we use the simplified decay width $\Gamma
(\omega)=g_{n}^{2}\omega^{5}\,\,\,\,$(valid for$\,\,\omega\lesssim
\omega_{\text{on-shell}}+m_{\pi}\,;$ for the full formula, see Ref.
\cite{berezhiani}). The on-shell values are $\omega_{\text{onshell}}%
=m_{n}-m_{p}-m_{e}=0.782333$ MeV and $\Gamma_{\text{onshell}}=g_{n}^{2}%
\omega_{\text{onshell}}^{5}=\hslash/\tau_{\text{n}}^{\text{beam}}%
=\hslash/888.1$ sec$^{-1}=7.41146\cdot10^{-25}$ MeV (out of which
$g_{n}=1.59028\cdot10^{-12}$ MeV$^{-2}$).

The fact that the function $\Gamma(\omega)$ is rising around $\omega
_{\text{onshell }}$implies that the IZE is possible if the time interval for
subsequent measurement $\tau$ is sufficiently short. In fact, $\Gamma
_{k}^{\text{meas}}(\tau,\omega_{C})>\Gamma_{\text{onshell}}$ for any value of
$\tau$ (and by using reasonable values of the off-shellness $\omega_{C}$; in
the following we shall use $\omega_{C}=5\omega_{\text{onshell}}$).%

\begin{figure}
[ptb]
\begin{center}
\includegraphics[
height=1.657in,
width=5.0323in
]%
{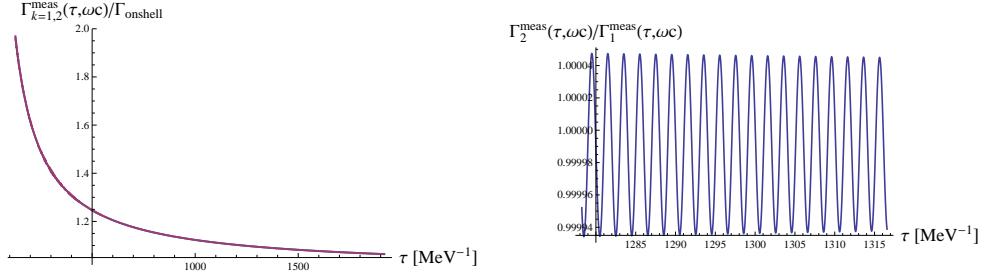}%
\caption{Left panel: the functions $\Gamma_{1,2}^{\text{meas}}(\tau,\omega
_{C}=5\omega_{\text{onshell}})/\Gamma_{\text{onshell}}$ are plotted as
function of $\tau.$ They are larger than unity, hence the IZE occurs. Both
curves are so similar that it is almost impossible to distinguish them. Right
panel: ratio $\Gamma_{2}(\tau,\omega_{C}=5\omega_{\text{onshell}})/\Gamma
_{1}(\tau,\omega_{C}=5\omega_{\text{onshell}})$ as function of $\tau$ in a
small time interval: the ratio is very close to one. This is true for other
values of $\tau$ as well.}%
\end{center}
\end{figure}

Unfortunately, we do not know which is the correct model for measurement in
the case of trap experiments.\ In\ Ref. \cite{izen} the function $f_{2}$ was
used for illustrative purposes. Here,we compare both models mentioned above to
see to which extent the results are comparable. Note, for both response
functions the numerical value of the `highest energy' $\omega_{C}$ is an
important parameter, since the measured decay width would diverge without this
cutoff. The value of $\tau$ for which the width in the trap experiment is
$\Gamma_{k}^{\text{meas}}(\tau,\omega_{C})/\Gamma_{\text{onshell}}=1.0098$ is
$\tau=12569.4$ MeV$^{-1}$ when using $f_{1}$ and $\tau=12569.9$ MeV$^{-1}$
when using $f_{2}$; these two values are extremely similar (for the discussion
about why this value of $\tau$ is reasonable for trap experiments, see Ref.
\cite{izen}). In Fig. 1 we also show the dependence on $\tau$ (we use slightly
smaller values of $\tau$ to see better the effect). The left panel shows that
the IZE occurs and that both response functions generate very similar results.
This is confirmed by the right panel, where the ratio is shown to be very
close to unity. Varying $\omega_{C}$ does not change the outcome as long as,
of course, the same value is used for both models.

\textit{Concluding remarks:} In this work we have tested two different models
of measurement that lead to IZE effect for the decay of the neutron in trap
experiments. Both of them lead to very similar results, thus showing that the
IZE is not dependent on the details of the measurement process. In the future,
one should repeat the previous study by using more advanced measurement models
which go beyond the simple bang-bang measurement (leading to $f_{1}$) or the
continuous measurement of the final state (leading to $f_{2}).$ It would be
particularly interesting to model the continuous measurement of the initial state.

\textbf{Acknowledgments}: The author thanks G. Pagliara for collaboration
leading to Ref. \cite{izen} as well as S. Mr\'{o}wczy\'{n}ski and P. Moskal
for useful discussions.

\end{document}